\documentclass[final,3p,times]{elsarticle}
\usepackage[utf8]{inputenc}
\usepackage{amssymb,amsmath,graphicx}
\usepackage{scrextend}
\usepackage{booktabs}
\usepackage[colorlinks=true,linkcolor=black,citecolor=blue,urlcolor=blue]{hyperref}
\newcommand{\eref}[1]{Eq.~(\ref{#1})}
\newcommand{\eg}{\emph{e.g.}}
\newcommand{\abs}[1]{\lvert #1\rvert}
\biboptions{authoryear,round}
\usepackage{lineno}
\journal{Journal of Informetrics}
\begin{document}
\begin{frontmatter}
\title{Unbiased evaluation of ranking metrics reveals consistent performance in science and technology citation data}
\author[label1]{Shuqi Xu}
\author[label1,label2]{Manuel Sebastian Mariani}
\author[label1,label3]{Linyuan Lü}
\ead{linyuan.lv@uestc.edu.cn}
\author[label1,label4,label5]{Matúš Medo}
\ead{matus.medo@unifr.ch}
\address[label1]{Institute of Fundamental and Frontier Sciences, University of Electronic Science and Technology of China, Chengdu 610054, PR China}
\address[label2]{URPP Social Networks, Universitat Zurich, 8050 Zurich, Switzerland}
\address[label3]{Alibaba Research Center for Complexity Sciences, Hangzhou Normal University, 311121 Hangzhou, PR China}
\address[label4]{Department of Radiation Oncology, Inselspital, Bern University Hospital and University of Bern, 3010 Bern, Switzerland}
\address[label5]{Department of Physics, University of Fribourg, 1700 Fribourg, Switzerland}

\begin{abstract}
Despite the increasing use of citation-based metrics for research evaluation purposes, we do not know yet which metrics best deliver on their promise to gauge the significance of a scientific paper or a patent. We assess 17 network-based metrics by their ability to identify milestone papers and patents in three large citation datasets. We find that traditional information-retrieval evaluation metrics are strongly affected by the interplay between the age distribution of the milestone items and age biases of the evaluated metrics. Outcomes of these metrics are therefore not representative of the metrics' ranking ability. We argue in favor of a modified evaluation procedure that explicitly penalizes biased metrics and allows us to reveal metrics' performance patterns that are consistent across the datasets. PageRank and LeaderRank turn out to be the best-performing ranking metrics when their age bias is suppressed by a simple transformation of the scores that they produce, whereas other popular metrics, including citation count, HITS and Collective Influence, produce significantly worse ranking results.
\end{abstract}

\end{frontmatter}

\section{Introduction}
Citation-based metrics for impact build on the premise that the number of citations received by a scientific paper (or a patent) is a reliable proxy for its scientific (or technological) impact. Such metrics are used not only to assess the impact of individual papers, but also to evaluate the overall research output of research units such as individual researchers~\citep{hirsch2005index,radicchi2009diffusion,zhou2012quantifying,medo2016model}, research institutes~\citep{charlton2007evaluating,west2013author}, and journals~\citep{harzing2009google,gonzalez2010new}, for example. The relative ease with which new metrics of research impact can be designed has contributed to their proliferation\citep{mingers2015review,waltman2016review,todeschini2016handbook}, and uncritical use of such metrics has eventually met a strong opposition \citep{manifesto2015,rijcke2016evaluation,leydesdorff2018h}. 
In particular, scholars have emphasized the need for understanding the theoretical foundations of impact metrics~\citep{waltman2016review}, and evaluating analytically and empirically the merits of the metrics~\citep{leydesdorff2018h}.

Despite the use of citation-based metrics for research evaluation purposes and the increasingly recognized need to better grasp their merits and pitfalls, we do not know yet which metrics best deliver on their promise to gauge the significance of a scientific paper or a patent. This gap is potentially dangerous: If a paper-level metric is assumed to be a good proxy for significance and it is used for research evaluation purposes, yet it undervalues papers whose significance is undeniable, its normative use~\citep{leydesdorff2018h} might lead to decisions that penalize truly significant research.

We aim to fill this gap by providing a comprehensive empirical comparison of a broad range of ranking metrics. We assess the metrics' ability to single out scientific papers and patents that have been recognized by field experts as groundbreaking or seminal. The core idea behind this evaluation is that metrics that aim to gauge the significance of a paper/patent should be able to detect papers/patents whose outstanding long-term significance for the involved fields is undeniable. Our goal is to answer, whether some metrics perform well across different citation datasets. If that is not the case, which characteristics of the input data decide which metric is the most suitable?

To this end, we analyze three citation networks: the scholarly citation dataset that includes papers published by the American Physical Society (APS), the citation data from the High-Energy Physics Literature Database INSPIRE (HEP), and the U.S. Patent Office patent citation data (PAT). We use expert-selected sets of seminal nodes and assess ranking metrics by how well are the seminal nodes ranked by them. In particular, we use milestone papers selected by APS journal editors for the APS data, community-curated ``Chronology of Milestone Events in Particle Physics'' for the HEP data, and the list of significant patents by \citet{strumsky2015identifying} for the PAT data (see Data section for details). While by no means exhaustive, these lists of seminal papers consist of papers and patents of exceptional importance (many papers have, for example, led to a Nobel Prize to one or more of their authors). Our evaluation includes nine network-based ranking metrics from the scientometrics and network science literature together with their time-normalized~\citep{mariani2016identification} variants. To provide a comprehensive comparison of metrics, we have chosen network-based metrics that have been used in bibliometrics~\citep{waltman2016review} or they have performed well in other networks (such as social and technological networks). Additionally, we provide results also for the percentile-based citation count which is commonly used in bibliometrics~\cite{leydesdorff2011turning,bornmann2013use}.

Expert-selected nodes have been used before to evaluate rankings of authors \citep{radicchi2009diffusion,dunaiski2019globalised}, rankings of movies \citep{wasserman2015cross,ren2018randomizing}, rankings of scientific papers \citep{mariani2016identification,dunaiski2019interplay}, and rankings of court cases~\citep{fowler2008authority}, for example (see \citep{dunaiski2018evaluate} for a recent in-depth discussion of this evaluation approach). We make here an important methodological distinction by distinguishing two similar, yet fundamentally different ranking tasks:
\begin{labeling}{Task 2}
\item[Task 1] The usual task for a ranking metric is to rank the given seminal nodes as high as possible. This is motivated by the assumption that if seminal nodes are known to be of high impact, a good metric should place them in the ranking of all nodes as high as possible. To evaluate the metrics' performance in this task, one typically uses traditional information-retrieval metrics such as precision, recall, and accuracy~\citep{dunaiski2016evaluating}.
\item[Task 2] The need for unbiased evaluation, which is common in scientometrics \citep{mutz2012generalized,mingers2015review,vaccario2017quantifying}, motivates the second task: To rank the expert-selected nodes as high as possible whilst requiring that the ranking metric is unbiased. Since most structural ranking metrics are biased (see \ref{sec:bias} for a demonstration), their evaluation must include a penalty for the performance gained thanks to the methods' bias.
\end{labeling}
Citation data commonly feature various biases such as the field bias~\citep{vaccario2017quantifying}, for example. We focus here on the \emph{age bias}; we say that age bias is present in the data when there is a dependence between the mean citation count and the publication age. Besides being particularly strong, the age bias is also explicit and easy to measure as it is determined by the paper's publication date. It thus provides a good test bed for dealing with biases in the ranking problem.

As part of our goal to evaluate the ranking performance of the chosen ranking metrics, we aim to elucidate the difference between the two ranking tasks described above. In particular, we show that Task 1 can favor biased metrics to such extent that a custom-constructed ranking method based only on the bias itself (in our case, ranking by age) can in some cases outperform all standard ranking metrics. The fact that a metric performs well in the common Task 1 is thus no direct indication of its ability to assess the value or impact of the network's nodes. We demonstrate that the normalized identification rate introduced in \citep{mariani2016identification} appropriately addresses Task 2 by imposing performance penalization proportional to the bias magnitude of the evaluated metric. Unlike Task 1, the results in Task 2 reveal consistent patterns of metrics' performance across the three studied datasets. As further discussed in Section~\ref{sec:from1to2}, the proposed evaluation can be also interpreted as a ranking problem where a given set of seminal nodes is interpreted as a potentially biased sample from a larger group of high-quality nodes. This further increases the relevance of Task 2 as compared to the common and straightforward approach presented in Task 1.

The paper is organized as follows. In Section~\ref{sec:data}, we describe and analyze the datasets and the corresponding sets of seminal nodes. In Section~\ref{sec:methods}, we present the considered network metrics and describe various performance measures of a metric's ranking ability based on seminal nodes. In Section~\ref{sec:task1}, we first address Task~1, evaluate the metrics by how well do they rank the seminal nodes, and eventually discuss drawbacks of this evaluation approach. In Section~\ref{sec:task2}, we address Task~2 where the metrics' bias is taken into account, and explain why are thus-obtained results more relevant than the results obtained within Task~1. Finally in Section~\ref{sec:discussion}, we compare the obtained results, discuss the relevance and scope of the two studied tasks, and review open research directions.

\section{Literature review}

\subsection{Citation-based ranking metrics}
Citation impact indicators are a key tool in scientometrics and play a prominent role in the evaluation of scientific and technological publications~\citep{waltman2016review}. The growing demand of evaluation information from researchers, funding bodies, and research institutions and the increasing availability of extensive data on scholarly activity have driven the proliferation of new indicators~\citep{mattedi2017evaluation}.

Among citation-based impact indicators for research articles, citation count (referred as indegree in the field of network science \citep{newman2010networks,zeng2017science}) is the most basic and established one as it has been used for ranking of scholarly publications since the 1970s \citep{liao2017ranking}. The basic premise citation count is simple: the most influential publications are the most cited. The metric's simplicity comes at a cost as the natural differences between citations are neglected by citation count (see \citep{bornmann2008citation} for an extensive review of the citing behavior). In particular, citation count assigns the same weights to a citation from a ground-breaking article published in a leading journal and a citation from an obscure article. The seminal PageRank algorithm for the World Wide Web \citep{brin1998anatomy} assigns higher weight to references from webpages that are highly valued by the algorithm. \cite{chen2007finding} applied PageRank to a citation network to measure the importance of individual scientific publications, initiating the interest in recursive citation impact indicators \citep{waltman2016review}. Since then, various PageRank variants \citep{waltman2014pagerank} have been proposed, of which the CiteRank \citep{walker2007ranking} is the best known. Of note is the HITS (Hyperlink-Induced Topic Search) algorithm \citep{kleinberg1999authoritative} which assigns two scores, hub score and authority score, to each node. In the context of citation data, it is natural to consider vital reviews as hubs which cite other influential publications that have high authority score and high-impact articles which tend to be cited by, among others, review articles with high hub scores \citep{nickerson2018measuring}. Another notable branch of research impact indicators is the $h$-index \citep{hirsch2005index} which was introduced to evaluate the scientific output of researchers and later extended by \cite{schubert2008using} to assess the impact of individual publications. A large number of variants of the $h$-index have been proposed in the literature \citep{alonso2009h}.

Besides the most used citation impact indicators, including citation count, PageRank, CiteRank, H-index and HITS authority score, we consider also several network-based metrics that performs well in other ranking scenarios: LeaderRank, Collective Influence, and Semi-local centrality. These metrics are introduced in Section~\ref{sec:methods}.

\subsection{Ranking bias in citation analysis}
In the analysis of citation data, citation counts of different publications cannot be directly compared as there are various sources of bias that can invalidate the validity of such a comparison. The most-studied biases are induced by the publication research field, age and document type \citep{waltman2016review}. Field bias manifests itself by the mean citation count of publications of a similar age varying greatly between fields because of their different citation practices \citep{lundberg2007lifting}. The average citation count differs between, for example, natural sciences and humanities by the factor of ten~\citep{bornmann2015methods}. The age bias of citation count has two distinct components. First, the average citation count of papers of a fixed age gradually increases over time~\citep{martin2013coauthorship}. Second, the citation counts naturally grow with the age of publications as they accumulate more and more citations with time which prevents citation counts of papers of different age from being directly comparable. Finally, citation counts of publications of different document types (such as article, letter, review, and commentary) should not be directly compared with each other because, for example, review papers tend to attract more citations than ordinary research articles and letters \citep{lundberg2007lifting}.

To overcome these biases, a number of normalized citation-based ranking metrics have been developed to allow for more fair comparisons, such as mean-based metrics and percentile-based metrics (see \cite{waltman2016review,bornmann2015methods} for reviews). In the case of paper and patent citation networks, that are our focus in this manuscript, the most relevant bias is the age bias which is strong and relevant in both paper and patent citation data.

Despite many citation-based ranking metrics being proposed in the past, a comprehensive comparison of their ranking performance on various datasets is still lacking. To determine which metrics are best suited to rank papers in citation networks is the first research gap that we aim to fill by this study. The common approach to assess a metric's ranking performance is based on using expert-selected nodes. The previous studies adopting this approach focus solely on the ranking positions of the expert-selected nodes and ignore the confounding effects that can be introduced by the choice of the seminal nodes. For example, if experts tend to identify old works as seminal, a ranking metric that shares this bias gains an advtangea and its potential superiority may be illusive. We fill this second research gap by exploring the interplay between the bias of the evaluated ranking metrics and the bias of the seminal nodes, and exploring an evaluation procedure that takes the bias of the seminal nodes into account.

\begin{table}
\centering
\begin{tabular}{rrrrrr}
\toprule
   Dataset & Label &  Time span &       $N$ &        $E$ & $S$\\
\midrule
APS papers &   APS & 1893--2016 &   595,287 &  7,051,801 & 160\\
HEP papers &   HEP & 1764--2017 &   829,708 & 14,994,123 & 310\\
US patents &   PAT & 1926--2010 & 6,237,625 & 45,962,301 & 112\\
\bottomrule
\end{tabular}
\caption{Basic characteristics of the networks corresponding to the three analyzed datasets: Their time span, the number of nodes $N$, the number of edges $E$, and the number of seminal nodes $S$.}
\label{tab:datasets}
\end{table}

\section{Data}
\label{sec:data}
To compare the ranking performance of network-based metrics, we use three citation datasets: the classical American Physical Society citation data, high-energy physics citation data, and the U.S. Patent Office citation data. Each dataset can be represented as a growing directed network where nodes gradually appear with time. Time resolution is one day for all datasets. Nodes represent papers or patents and directed links represent citations. For each dataset, there is a set of corresponding expert-selected nodes of high impact that we refer as seminal nodes. Table~\ref{tab:datasets} summarizes basic characteristics of the analyzed networks and the corresponding sets of seminal nodes.

\subsection{American Physical Society citation data (APS)}
The American Physical Society (APS) dataset in our possession covers years 1893--2016 (the dataset is available on demand from \url{https://journals.aps.org/datasets}). After removing non-research papers (announcements, book reviews, etc.), the dataset contains $595,287$ nodes (papers) published by the APS journals and $7,051,801$ directed links (citations) between them. For this dataset, we use multiple selections of milestone papers chosen by editors of APS journals: 87 Physical Review Letters milestones\footnote{Retrieved from \url{https://journals.aps.org/prl/50years/milestones} on June 6, 2017.},
23 Physical Review E milestones\footnote{Retrieved from \url{https://journals.aps.org/pre/collections/pre-milestones} on June 6, 2017.},
and 78 select papers announced on the 125th anniversary of the Physical Review journals.\footnote{Retrieved from \url{https://journals.aps.org/125years} on January 12, 2018.}
In total, there are 161 unique seminal papers, of which 160 are present in the citation data (the one missing paper is from 2017, hence outside the coverage period of our dataset).

\subsection{High-Energy Physics citation network (HEP)}
INSPIRE is a project run by leading high-energy physics institutions around the world (CERN, DESY, Fermilab, IHEP, and SLAC). Among other things, it curates a database of papers high-energy physics papers (and papers relevant to the high-energy physics community), which is also made available on demand for research purposes.\footnote{We downloaded the INSPIRE data on October 30, 2017 from \url{https://inspirehep.net/dumps/inspire-dump.html}.}
After processing the downloaded xml data dump covering years 1764--2017, we obtained a citation network containing $829,708$ nodes and $14,994,123$ directed links. The list of milestone papers has been downloaded from the website ``Chronology of Milestone Events in Particle Physics''\footnote{Retrieved from \url{http://web.ihep.su/dbserv/compas/} on April 6, 2018.}
that lists milestone events and the corresponding papers. The website is a joint effort of the Institute for High Energy Physics (Russia) and the Particle Data Group (USA) with several leading high-energy researchers contributing to the final version of the chronology. Thus-obtained milestone papers have been matched with the HEP data, leading to the final set of 310 seminal papers.

\subsection{US patent citation network (PAT)}
The US patent dataset was collected by \cite{kogan2017technological} and covers years 1926--2010. In total, there are $6,237,625$ nodes (patents) and $45,962,301$ links (citations) among them. In \cite{strumsky2015identifying}, the authors listed 175 patents which ``affected society, individuals and the economy in a historically significant manner''. In agreement with \cite{mariani2018early}, we remove the patents issued outside the citation dataset's time span as well as the design patents that are absent in the citation data. As a result, $112$ seminal patents are used for further analysis.

\begin{table}
\centering
\begin{tabular}{rrrrr}
\toprule
Dataset    & Set of nodes & Median indegree & $\tau_3$ & $\tau_5$\\
\midrule
APS papers &          All &   5 &  3.6 years &  4.8 years\\
           &      Seminal & 239 &  1.3 years &  2.2 years\\
\midrule
HEP papers &          All &   4 &  3.1 years &  3.9 years\\
           &      Seminal &  88 &  7.6 years & 12.4 years\\
\midrule
US patents &          All &   3 & 11.5 years & 13.3 years\\
           &      Seminal &  30 &  9.6 years & 12.0 years\\
\bottomrule
\end{tabular}
\caption{A comparison between the seminal nodes and all nodes. Here $\tau_3$ and $\tau_5$ are the mean times needed for the nodes to get their first 3 and 5 citations, respectively (ignoring the nodes that have less than 3 and 5 citations, respectively).}
\label{tab:all_vs_seminal}
\end{table}

Table~\ref{tab:all_vs_seminal} compares the characteristics of all nodes and the seminal nodes in each dataset. As expected, the seminal nodes have indegree (commonly referred to as the number of citations) significantly higher than the overall median indegree in all three datasets. This confirms that the expert assessment of the seminal nodes is not in contradiction with their impact as reflected by the citation network.

Table~\ref{tab:all_vs_seminal} further lists the times needed to collect their first three and five citations respectively, $\tau_3$ and $\tau_5$, by various groups of nodes. These times indicate the timescales of citation dynamics. We see, for example, that the nodes collect their citations in the PAT data significantly slower than in the APS and the HEP data. In the APS data, both $\tau_3$ and $\tau_5$ are smaller for the seminal nodes than they are for all nodes, which is understandable given the much higher indegree of the seminal nodes. Curiously, the relation is reverted for the HEP data where the seminal nodes need more time to get their first 3 or 5 citations than all papers. While paradoxical at first sight, this is a direct consequence of the HEP seminal nodes being over-represented among the old nodes (see Figure~\ref{fig:age_distribution}). At the time when these seminal nodes were collecting their first citations, the citation dynamics was substantially slower than nowadays, and this then manifests itself in their high $\tau_3$ and $\tau_5$. We shall see in Section~\ref{sec:from1to2} that the strong age bias of the HEP seminal nodes has further important implications.

\section{Node ranking metrics}
\label{sec:methods}
We use nine distinct network centrality metrics that are described below (see Table \ref{tab:allMetrics} for a summary), and their variants where the age bias of metrics has been removed by the rescaling procedure introduced in \cite{mariani2016identification} (see \cite{vaccario2017quantifying} for simultaneous removal of age and field bias by the rescaling procedure). In a dataset with time-stamped nodes, rescaling can be applied to any node ranking metric; see Section~\ref{sec:rescaling} for details. The input citation data are represented by the $N\times N$ adjacency matrix $\mathsf{A}$ whose element $A_{ij}=1$ if node $i$ cites node $j$ and $A_{ij}=0$ otherwise.

\subsection{Citation count (C)}
Citation count refers to the number of citations received by a given paper or patent. It is equivalent to node indegree in a directed network \citep{newman2010networks}. For node $i$, citation count is defined as $C_i=\sum_{j=1}A_{ji}$. Based on the assumption that a node is important if it is cited by many other nodes, citation count is the simplest and the most widely used indicator of paper or patent impact in citation data. The metric's simplicity directly translates into its low computational complexity.

\subsection{PageRank (P)}
PageRank \citep{brin1998anatomy}, which is an node centrality metric originally devised to rank pages in the World Wide Web and later applied to citation data to assess the significance of publications \citep{chen2007finding}, introduces the importance of different nodes in a self-consistent manner: A node is important if it is cited by other important nodes. PageRank score $P_i$ of node $i$ is defined by the set of equations \citep{berkhin2005survey}
\begin{equation}
\label{PR}
P_i=\alpha\sum_{j:k^{out}>0}\frac{A_{ji}}{k_j^{out}}\,P_j+
\alpha\sum_{j:k^{out}=0}\frac{P_j}{N}+\frac{1-\alpha}{N}
\end{equation}
where $i=1,\dots,N$, $k_j^{out}=\sum_{l}A_{jl}$ is the outdegree of node $j$, $\alpha$ is the damping factor, and $(1-\alpha)/N$ is usually referred to as the teleportation term whose role is to ensure that \eref{PR} has a unique solution. While $\alpha=0.85$ is used in the ranking of web pages \citep{berkhin2005survey}, $\alpha=0.5$ is typically used in the analysis of citation data \citep{chen2007finding}. \eref{PR} is usually solved by iterations: Starting from the uniform initial score $P_i^{(0)}=1/N$, every node's score is updated iteratively as \citep{berkhin2005survey}
\begin{equation}
\label{PR2}
P_i^{(n+1)}=\alpha\sum_{j:k^{out}>0}\frac{A_{ji}}{k_j^{out}}\,P_j^{(n)}+
\alpha\sum_{j:k^{out}=0}\frac{P_j^{(n)}}{N}+\frac{1-\alpha}{N}
\end{equation}
where $n$ is the iteration number. We stop the iterations when the average score change is small enough, that is $\sum_{i = 1}^N\abs{P_i^{(n)}-P_i^{(n-1)}}/N<\varepsilon$ where $\varepsilon=10^{-9}$. The same stopping condition is used also in other metrics that involve iterations (CiteRank and LeaderRank).

\subsection{CiteRank (T)}
CiteRank \citep{walker2007ranking} has been introduced to offset PageRank's strong bias towards old nodes [note that in some cases, PageRank can be also biased towards recent nodes \citep{mariani2015ranking}]. Using the representation of PageRank as a random walk on the citation network, CiteRank modifies the algorithm by initially distributing random walkers preferentially on recent nodes, with the old nodes being exponentially suppressed at a timescale $\tau$. Similarly to \eref{PR2}, CiteRank score $T_i$ for node $i$ can be defined in an iterative way as
\begin{equation}
T_i^{(n+1)}=\alpha\sum_{j:k^{out}>0}\frac{A_{ji}}{k_j^{out}}\,T_j^{(n)}+
\alpha\sum_{j:k^{out}=0}\frac{T_j^{(n)}}{N}+(1-\alpha)\,
\frac{\exp\big[-(t-t_i)/\tau\big]}{\sum_{j=1}^{N}\exp\big[-(t-t_i)/\tau\big]}
\end{equation}
where $t_i$ is the publication date of node $i$, $t$ is the date when the scores are computed; other terms and parameters have the same meaning as for PageRank. To estimate suitable values for parameters $\alpha$ and $\tau$, we followed the procedure described in \citet{walker2007ranking} where the authors maximize the correlation between the CiteRank scores and the nodes' recent indegree increase. The resulting parameter values are $\alpha=0.50$, $\tau=2.6\,\text{years}$ for APS; $\alpha=0.50$, $\tau=2.4\,\text{years}$ for HEP; and $\alpha=0.44$, $\tau=7.6\,\text{years}$ for PAT. Notably, the parameter values for APS are the same as reported in \citet{walker2007ranking} despite our APS dataset including 13 additional years and, consequently, 60\% more papers.

\subsection{LeaderRank (L)}
The need for a teleportation term in PageRank can be eliminated by connecting each node to an artificial ``ground'' node with bidirectional links. The resulting parameter-free LeaderRank metric has been proposed in \citep{lu2011leaders} to quantify node influence. The iterative equation for the LeaderRank score $L$ is
\begin{equation}
L_i^{(n+1)}=\sum_{j}^{N+1}\frac{A_{ji}}{k_j^{out}}L_j^{(n)}
\end{equation}
where both $A_{ji}$ and $k_j^{out}$ include the ground node and the links between the ground node and all other nodes in the network. After obtaining the equilibrium scores $L_i^{(n_c)}$, the ground node is removed from the system and its score is evenly distributed among all real nodes. The final score of node $i$ is thus defined as
\begin{equation}
L_i=L_i^{(n_c)}+\frac{L_g^{(n_c)}}{N}.
\end{equation}
The redistribution of the ground node's score does not affect the ranking of nodes by LeaderRank, though.

\subsection{H-index (H)}
$H$-index \citep{hirsch2005index} was originally devised to characterize the academic impact of researchers based on their publications and citations \citep{hirsch2005index,hirsch2007does}. Similarly as it was later applied to evaluate research journals \citep{braun2006hirsch}, it can be adapted also to evaluate research papers: The $h$-index of paper $i$ is defined as the largest number $h_i$ such that paper $i$ is cited by at least $h_i$ papers that each have at least $h_i$ citations \citep{schubert2008using,lu2016h}.

\subsection{Collective Influence (CI)}
Collective Influence was introduced in \citet{Morone2015Influence} to identify those nodes that, when removed, cause the biggest damage to a graph's giant component; the algorithm is based on the classical problem of percolation in complex networks. The $CI$ centrality of node $i$ at level $l$ is defined as
\begin{equation}
CI_i^l=(k_i-1)\sum_{j: d_{ij}=l}(k_j-1)
\end{equation}
where $k_i$ is the degree of node $i$, $d_{ij}$ is length of the shortest distance between nodes $i$ and $j$, and $l$ is the metric's parameter. In line with \cite{bovet2019influence}, we consider only node indegree in the computation of $CI$ as node indegree is indicative of node impact. Using node outdegree or combining in- and out-degree leads to inferior results in our evaluation. Distance $d_{ij}$ is computed so that it respects link directions. Our tests show that $l=1$ and $l=2$ produce the best results, the metric's performance deteriorates as $l$ increases further. We use $l=2$ for all $CI$ results presented here.

\subsection{Semi-Local Centrality (SLC)}
Semi-local centrality was proposed \citep{chen2012identifying} as an extension of the purely local node degree (which is the simplest node centrality metric). It is semi-local in the sense of considering the node neighborhood up to the fourth order. The semi-local centrality score of node $i$ is defined as
\begin{equation}
SLC_i = \sum_{j\in\Gamma_i}Q_j,\qquad
Q_j = \sum_{k\in\Gamma_j}N_k
\end{equation}
where $\Gamma_i$ is the set of the nearest neighbors of node $i$ and $N_k$ is the number of the nearest and the next-nearest neighbors of node $k$. In this paper, we consider only the in-neighbors ($i$'s nearest in-neighbors are the nodes that cite node $i$) to leverage the impact of citations. If SLC is computed using out-neighbors or using both in- and out-neighbors, its performance (as measured by the metrics introduced in Sections~\ref{sec:IR} and \ref{sec:NIR}) deteriorates.

\subsection{Hyperlink-Induced Topic Search (HITS)}
HITS \citep{kleinberg1999authoritative} is a seminal ranking algorithm that considers two roles for each node in the network, authority and hub. A good authority is pointed by many hubs, and a good hub points to many authorities. The authority score of a node is equal to the sum of the hub scores of all nodes that point to this node, and the hub score of a node is equal to the sum of the authority scores of all nodes that this node points to. Mathematically, the authority score $a_i$ and the hub score $h_i$ of node $i$ fulfill
\begin{equation}
a_i^{n+1} = \sum_{j=1}^{N}A_{ji}h_j^{n},\qquad
h_i^{n+1} = \sum_{j=1}^{N}A_{ij}a_j^{n}.
\end{equation}
Both scores are normalized after each iteration so that the the sum over all nodes is one for each score. The iterations stop when the average score change is small enough, that is, $\sum_{i = 1}^N(\abs{a_i^{(n)}-a_i^{(n-1)}}+\abs{h_i^{(n)}-h_i^{(n-1)}})/N<\varepsilon$ where $\varepsilon=10^{-9}$. Of the two scores, authority is related to node impact in a citation network as it is derived from incoming links as opposed to hub which is derived from outgoing links (references) to nodes of high authority. We thus consider the equilibrium authority value as the nodes' HITS scores here.

\subsection{Yearly citation count percentile (YCCP)}
The use of percentiles in the ranking of papers has the advantage of avoiding working directly with citation counts that are typically broadly distributed which makes it difficult to aggregate them~\citep{leydesdorff2011turning} by, for example, averaging (such as computing the average citation count of the papers authored by an individual researcher). To reduce the age bias in the resulting ranking, we compute the citation count percentile of a node with respect to the citation counts of all nodes that have appeared in the same year as the target node. The nodes are finally ranked by their respective percentile ranks. Note that by comparing with nodes that appeared in the same year, this ranking metric already addresses the age bias; we thus do not consider a rescaled version of this metric. If the citation count percentile is computed with respect to all nodes regardless of their appearance time, the ranking is the same as the ranking by citation count, C.

\subsection{Rescaled metric variants}
\label{sec:rescaling}
To suppress the age bias of ranking metrics, we use the rescaling procedure proposed by \citet{mariani2016identification}; see \citep{dunaiski2019interplay} for other approaches to score normalization. The rescaled score $R(m_i)$ for metric $m$ and node $i$ is computed as
\begin{equation}
\label{eq:rescale}
R(m_i)=\frac{m_i-\mu_i(m)}{\sigma_i(m)}
\end{equation}
where $m_i$ is the original score of node $i$ as produced by metric $m$, and $\mu_i(m)$ and $\sigma_i(m)$ are the metric mean and standard deviation, respectively, computed over nodes in a window centered at node $i$. Assuming that the nodes are sorted by their age/appearance time, the window around node $i$ includes nodes $j\in[i - W/2, i + W/2]$ where the parameter $W$ represents the window size. For the APS, HEP, and PAT data, we use $W=1,000$, $W=2,000$, and $W=15,000$, respectively, which is roughly proportional to the number of nodes in each dataset.

As shown in \citet{mariani2016identification,mariani2018early,ren2019age}, rescaling significantly reduces the magnitude of the age bias---and, in the case of \cite{vaccario2017quantifying}, of the age and field bias---of citation count and PageRank. We use this technique here to rescale all ranking metrics introduced above, and in turn compare their performance with original non-rescaled metrics. Rescaled metrics are marked by adding $R$ at the beginning of their original labels (\eg, $RP$ for rescaled PageRank). The effectiveness of the rescaling procedure in removing the age bias of respective metrics in the studied datasets is investigated in \ref{sec:bias} where we find that rescaling indeed significantly reduces the age bias for almost all dataset-metric pairs (see Table~\ref{tab:bias_strength} for a summary).

\begin{table}
\centering
\begin{tabular}{rrr}
\toprule
Metric & Abbreviation & Implementation reference\\
\midrule
                  Citation count &    C & \cite{newman2010networks}\\
                        PageRank &    P & \cite{chen2007finding}\\
                        CiteRank &    T & \cite{walker2007ranking}\\
                      LeaderRank &    L & \cite{lu2011leaders}\\
                         H-index &    H & \cite{hirsch2005index}\\
   Directed collective influence &   CI & \cite{bovet2019influence}\\
           Semi-local centrality &  SLC & \cite{chen2012identifying}\\
                  HITS authority & HITS & \cite{kleinberg1999authoritative}\\
Yearly citation count percentile & YCCP & \cite{leydesdorff2011turning}\\
\bottomrule
\end{tabular}
\caption{The summary table of all metrics; our computation of each metric is based on the provided implementation reference. We included also age-normalized variants of the displayed metrics in the analysis (except for YCCP that already involves age-normalization). Except for the rescaled citation count~\cite{newman2009first} and the rescaled PageRank~\cite{mariani2016identification}, the rescaled variants of the remaining metrics have not been considered before.}
\label{tab:allMetrics}
\end{table}

\section{Metric performance in ranking the seminal nodes (Task 1)}
\label{sec:task1}
We first evaluate the ranking performance of metrics taking into account solely the ranking positions of the seminal nodes, in line with common information-retrieval practices \citep{radicchi2009diffusion,manning2010introduction,lu2011link,dunaiski2018evaluate}.

\subsection{Identification rate}
\label{sec:IR}
Our basic evaluation procedure is based on a complete given network which is used as an input. We rank the network nodes by their score according to a given metric $m$ and compute the fraction of the seminal nodes that are among the top $zN$ nodes, $f_z(m)$. This quantity is commonly referred as \emph{recall} in information filtering literature \citep{lu2012recommender}. To comply with previous research on rescaling \citep{mariani2016identification}, and also to avoid confusion for a related age-dependent version of this metric (see the next paragraph), we use here the previously-coined term \emph{identification rate} ($IR$) for $f_z(m)$. Note that $z\in(0, 1)$ is an evaluation parameter that, to reflect our goal of evaluating the ranking metrics by whether they rank the seminal nodes ``highly'', should be a small number. We use $z=1\%$ unless stated otherwise and later verify our main results using $z=0.5\%$ and $z=2\%$, respectively.

Besides assessing the identification rate on a complete network, we also study the metrics' performance as a function of the age of the seminal nodes \citep{mariani2016identification}. To this end, we construct network snapshots at the end of each calendar year (ignoring all nodes and links that appear afterward), and rank the nodes in each network snapshot. This allows us to evaluate, individually for each seminal node, whether it was at the top $z$ fraction of the ranking at any given age $\Delta t$. By averaging this over all seminal nodes\footnote{If a seminal node appears $t$ years before the end of the complete dataset, it is obviously impossible to know its ranking at age $\Delta t > t$. Seminal nodes that are younger than $\Delta t$ are therefore excluded from the averaging.}, we obtain the identification rate $f_z(m,\Delta t)$ which is now a function of the age of seminal nodes. $f_z(m, 1\,\text{year})$, for example, is the fraction of the seminal nodes that are in the top $z$ fraction of the ranking when they are one year old.

Task 1 focuses solely on the ranking positions of the seminal nodes, and these are reflected by $f_z(m)$ and $f_z(m, \Delta t)$. While the former evaluates the ``final'' ranking positions of the seminal nodes, the latter allows us to inspect how fast (or slow) do the seminal nodes rise in the rankings by the respective metric.

\begin{figure}
\centering
\includegraphics[width=\textwidth]{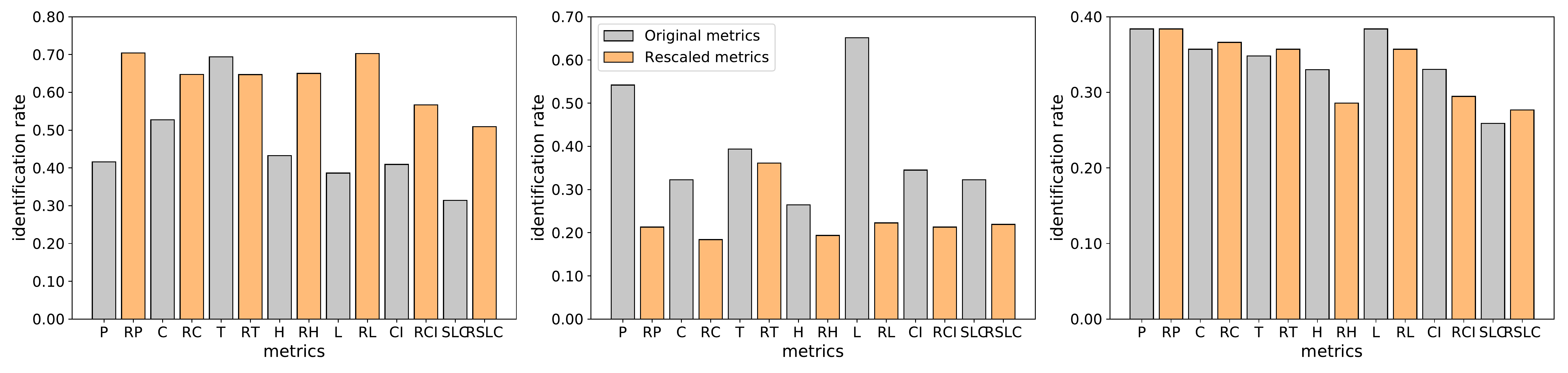}
\caption{Metrics' performance in identifying the seminal nodes as measured by the identification rate ($z=1\%$) in complete datasets. Note that the maximal displayed values differ between the panels. Colors of the bars are used to distinguish the original ranking metrics (grey) and their age-rescaled counterparts (orange).}
\label{fig:IR_result}
\end{figure}

\subsection{Metric evaluation using identification rate}
Figure~\ref{fig:IR_result} shows the ranking metrics evaluated by their $IR$ in complete datasets. Overall, the highest identification rates are found in APS, followed by HEP, and then by PAT. A likely reason for this is provided by Table~\ref{tab:all_vs_seminal} which shows that in the PAT data, median indegree differs the least between all nodes and the seminal nodes, thus making the seminal nodes in this dataset difficult to be separated from the other nodes.

The relative standings of metrics are rather similar between APS and PAT with PageRank being the best-performing metric in both. Relative differences between the metrics in both datasets are rather small, though: In both datasets there are a few methods with nearly-identical performance, and the ratio between the best and the worst metric's $IR$ is around 1.5. The results are very different in HEP where: (1) LeaderRank ($L$) outperforms the second-best method by a wide margin, (2) The ratio between the best and the worst metric's $IR$ is 3.5, (3) All rescaled metrics perform significantly worse than their unrescaled counterparts. We focus on metric evaluation in this section; reasons for the differences observed in HEP are discussed in Section~\ref{sec:from1to2}.

In summary, LeaderRank can be considered as the best-performing metric in Task 1 as it is clearly best in the HEP data and nearly-best in the APS and PAT data. This holds also when different evaluation thresholds, $z=0.5\%$ and $z=2\%$, are used.

\begin{figure}
\centering
\includegraphics[width=\textwidth]{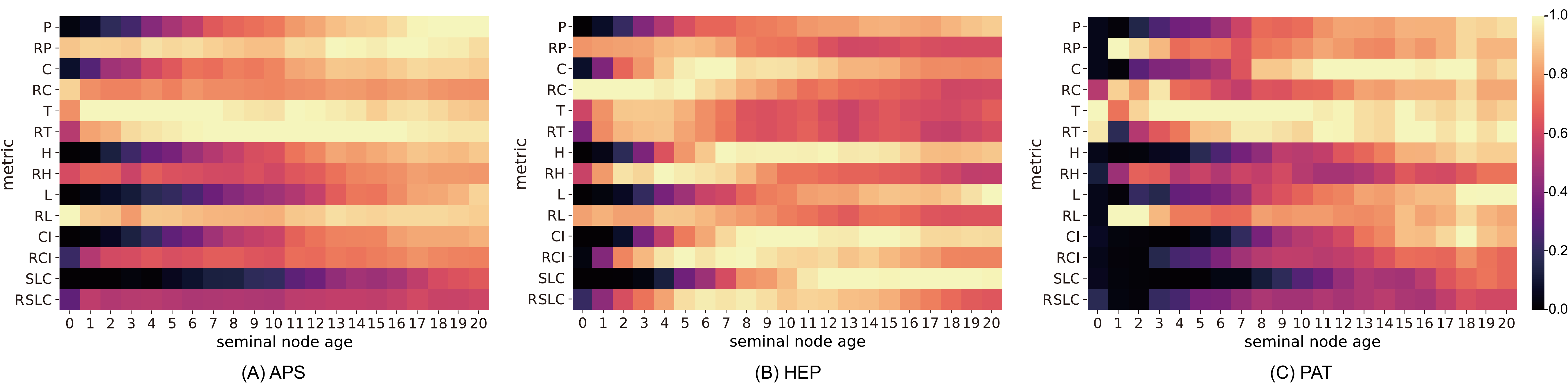}
\caption{The identification rate of individual metrics as a function of the seminal node age (in years). To facilitate the comparison, the metrics' performance is normalized to the best metric in each age bin. A metric with zero $IR$ thus receives zero score, while a metric that achieves the best $IR$ for given seminal node age receives the score of one.}
\label{fig:relative_result}
\end{figure}

Figure~\ref{fig:relative_result} consequently shows the metrics' relative performance as a function of the seminal node age. This approach serves to reveal the time evolution of the metrics' ranking performance. To further facilitate the comparison of metrics, we normalize the metrics' identification rate at a given age $\Delta t$ of seminal nodes, $f(m;\Delta t)$, by the best-achieved $IR$ at this age, $\max_n f(n;\Delta t)$. Relative performance thus ranges from zero (when a metric's $IR$ is zero at age $\Delta t$) to one (achieved by the best-performing metric at age $\Delta t$). As shown in Figure~\ref{fig:relative_result}, the relative performance of metrics changes dramatically with the seminal node age: metrics that work well short after publication (mostly rescaled metrics) lose their advantage as the seminal nodes become older. In the displayed node age range, CiteRank and rescaled CiteRank (T and RT) are two best-performing metrics in APS and PAT. For HEP, there is no single metric that performs well for most age values. Rescaled citation count (RC) is best until age 5, then $h$-index and collective influence (H and CI) are best until age 12, and finally semi-local centrality (SLC) is the best from then until age 20. LeaderRank (L), which performed best for the complete HEP  dataset in Figure~\ref{fig:IR_result}, becomes the best metric later on (for comparison, the average age of the seminal nodes in the complete HEP dataset is 61 years).

\begin{figure}
\centering
\includegraphics[width=\textwidth]{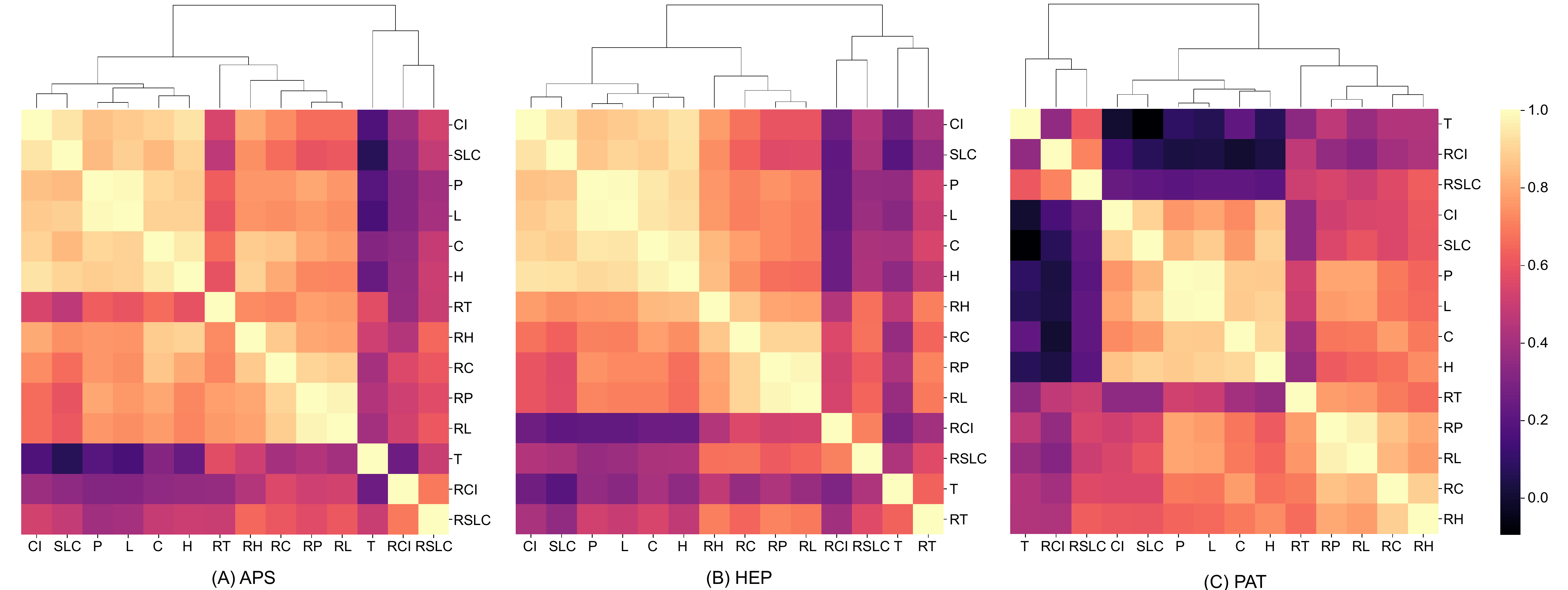}
\caption{Similarity of the evaluated metrics as measured by the Spearman rank correlation of all node rank positions. The metrics' hierarchical clusterings are obtained by the UPGMA method implemented by the clustermap function in Python's Seaborn library.}
\label{fig:spearman_similarity}
\end{figure}

To gain further insights in differences between the metrics, we evaluate their pairwise similarity using the Spearman rank correlation of all nodes' rankings. The results are shown in Figure~\ref{fig:spearman_similarity} together with metric clustering based on the obtained correlation matrices. There are several points to note. First, the clustering of metrics is remarkably stable across the datasets. Second, the clusterings reveal two groups of metrics whose rankings are similar to each other. The larger group includes CI, SLC, P, L, C, and H. The smaller group includes some of their rescaled variants: RP, RL, RC, and RH. Third, RCI, RSLC, and RT do not cluster with other rescaled metrics, probably as a result of the rescaling procedure not working perfectly for them (see Figures~\ref{SIfig:rankingbias_APS}, \ref{SIfig:rankingbias_HEP} and \ref{SIfig:rankingbias_PAT} in \ref{sec:bias}). Fourth, within each of the two mentioned clusters, the pairwise Spearman correlation values are rather high (above 0.73 in all three datasets), which indicates a high degree of similarity among the respective metrics.

Note that we have omitted HITS from the presentation of results above. The reason for doing so is that its performance is so much worse than that of the other metrics that the added value of displaying HITS in all previous figures would be very limited. In particular, the identification rate values of the HITS authority score are 0.143 (APS), 0.116 (HEP), and 0.054 (PAT).\footnote{HITS performance is strongly inferior to other metrics also in terms of normalized identification rate introduced in Section~\ref{sec:NIR}.}
The poor performance of HITS here is very different from this algorithm being praised in the line of research on court decision citation networks (see \cite{fowler2008authority,agnoloni2015case} and the references therein).
One possible reason for this difference is that in science, few would agree that a citation from a well-referenced but little cited review paper is more indicative of the target paper's impact than a citation from a high-impact paper with few references (as HITS authority score would assume). In this sense, court decision citation networks may be intrinsically more favorable to HITS than the paper and patent citation networks are. Further research is necessary to understand structural differences between court decision citation networks and scholarly/patent citation networks. Also, a comprehensive evaluation of several ranking metrics can help us understand whether HITS is indeed the singular best-performing metric in court decision networks.

We have similarly omitted yearly citation count percentile, YCCP, from the figures. Albeit the performance of YCCP does not lack behind the top metrics as much as the performance of HITS, the results are still significantly lower: The identification rate values of YCCP are 0.700 (APS), 0.197 (HEP), and 0.375 (PAT). Importantly, the $IR$ results of YCCP are similar to the results of RC which is expected as RC too, is an age-normalized version of citation count similarly to YCCP. Because of this high level of similarity, we report the YCCP results only in text.

\begin{figure}
\centering
\includegraphics[width=\textwidth]{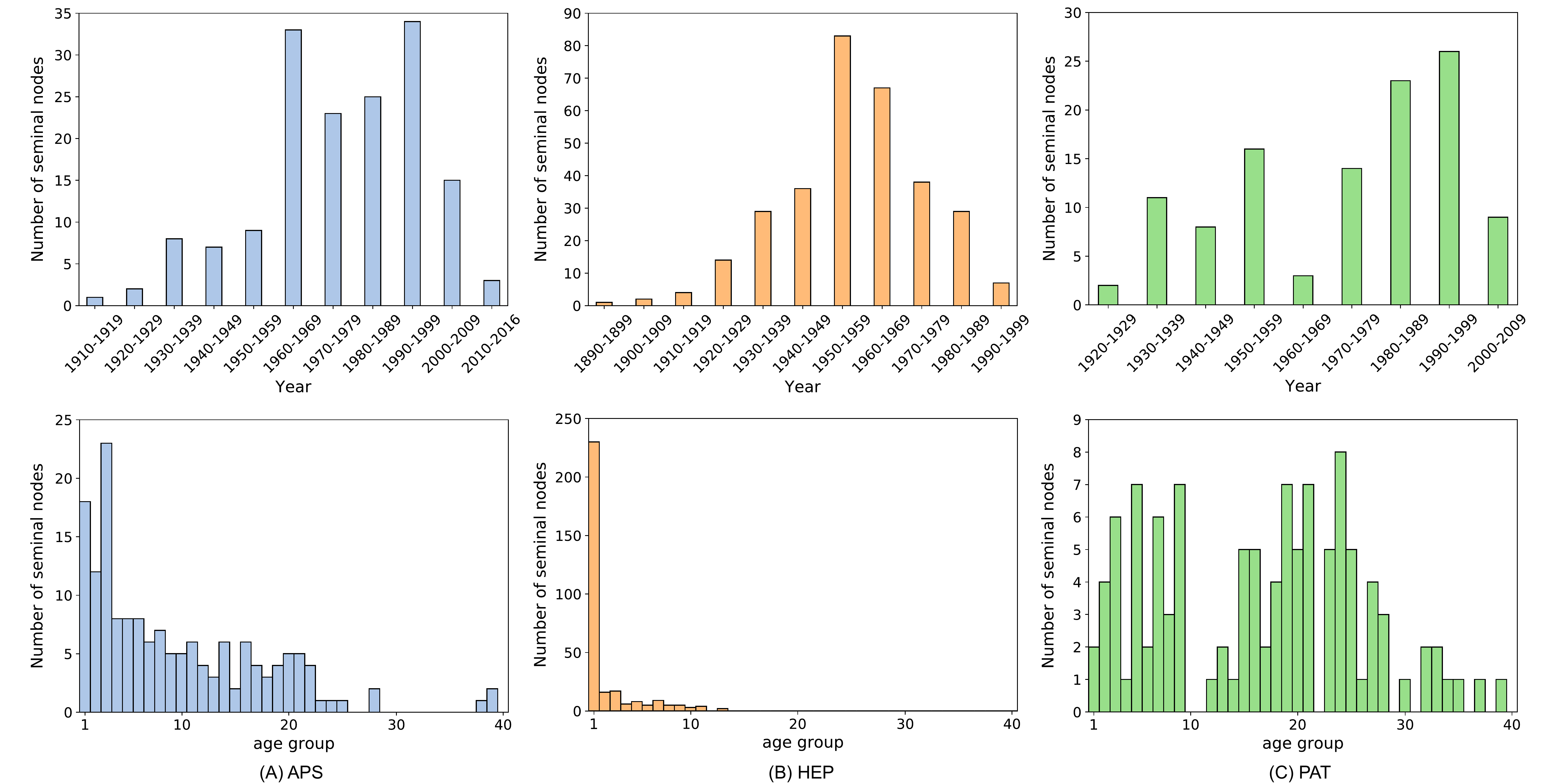}
\caption{The distributions of the seminal nodes' publication dates in the datasets: real time (top row), and 40 equally-sized age groups (bottom row).}
\label{fig:age_distribution}
\end{figure}

\subsection{Caveats of identification rate}
\label{sec:from1to2}
While metrics' performance in Figure~\ref{fig:IR_result} is strikingly uniform in the APS and PAT data, big differences are found in the HEP data. To explain where do they come from, Figure~\ref{fig:age_distribution} shows the age distributions of the seminal nodes in the data. In terms of real time, the difference between APS/PAT and HEP is already apparent as the first two datasets have the average publication year of the seminal nodes 1976 and 1975, respectively, whereas it is 1957 for the HEP seminal nodes. The difference between the three datasets is more evident, though, when each seminal node is assigned to one of the 40 equally-sized age groups by its publication date (with groups 1 and 40 containing the oldest and the most recent nodes, respectively). The bottom row of Figure~\ref{fig:age_distribution} shows that the HEP seminal nodes are distributed extremely unevenly among the age groups with 74\% of them (230 out of 310) in the oldest age group 1, and no seminal nodes in age groups 14--40. The big differences between the top and bottom row in Figure~\ref{fig:age_distribution} are due to the accelerating rates at which new nodes appear in the datasets. The numbers of recent new nodes are so high that they ``push'' the seminal nodes to the early age groups. In APS, for example, approximately 85\% of all nodes appear after 1976 which is the mean publication year of the dataset's seminal nodes.

\begin{figure}
\centering
\includegraphics[width=\textwidth]{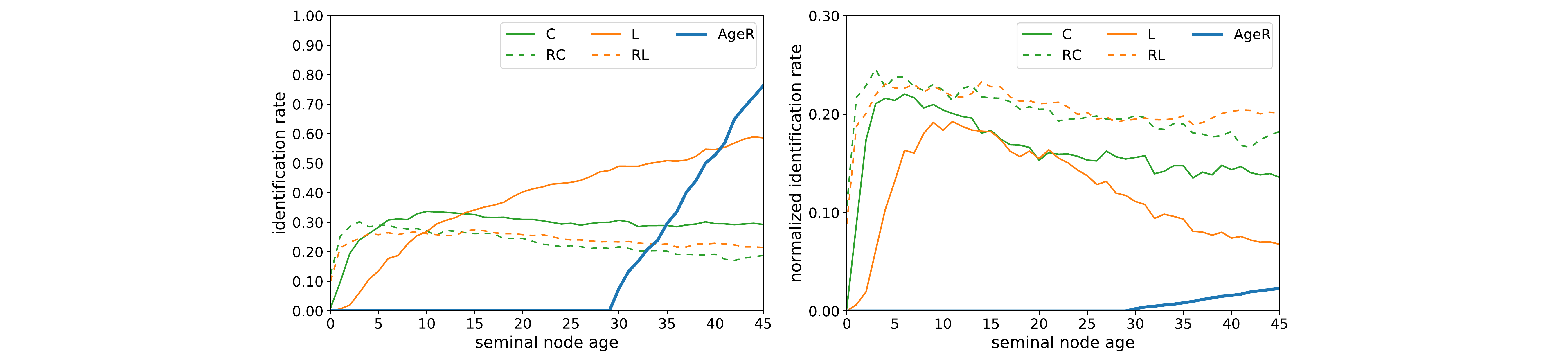}
\caption{Performance of selected metrics in identifying the seminal nodes in the HEP data: A comparison between the identification rate (left) and the normalized identification rate (right; see Section~\ref{sec:NIR} for the definition) measured as functions of the seminal node age.}
\label{fig:hep_IR_NIR}
\end{figure}

The strong temporal non-uniformity of the seminal nodes has profound consequences. Firstly, it is not favorable to age-rescaled metrics which, by design, strive for a uniform representation of all age groups among the top-ranked nodes. For the HEP data, however, nodes from age groups 2--40 can contribute only marginally to the identification rate because there are only a few seminal nodes among them. By contrast, original non-rescaled metrics are typically biased towards old nodes (see Figures~\ref{SIfig:rankingbias_APS}--\ref{SIfig:rankingbias_PAT}) and this gives them an advantage when a given set of seminal nodes shares the same bias towards old nodes. In particular, Figure~\ref{SIfig:rankingbias_HEP} shows that the bias of LeaderRank towards old nodes is the strongest of all metrics in HEP, which directly contributes to the metric's superior performance in Figure~\ref{fig:IR_result}.

Secondly, the age bias of the seminal nodes in HEP is so strong that it allows the simple ranking of nodes by their age (we refer this metric as AgeR; old nodes are at the top) to outperform all other metrics. Its identification rate on the complete HEP data is 0.70 which is indeed better than the values shown in Figure~\ref{fig:IR_result} for the other metrics. This is further illustrated by the left panel of Figure~\ref{fig:hep_IR_NIR} which shows the identification rate for a few selected metrics as a function of the seminal node age. Here AgeR yields zero identification rate when the seminal nodes are young (younger than 30 years) because it simply puts old nodes at the top of the ranking. However, the metric's results quickly improve when the seminal nodes are older than that and AgeR becomes the best metric starting from age 40, approximately. This demonstrates that evaluating ranking metrics solely by the ranks that they assign to the seminal nodes is of limited relevance as AgeR---a metric that entirely ignores the actual impact of the nodes---is eventually able to outperform all other metrics.

In summary, the identification rates observed within Task 1 are a confound outcome of a given metric's ability to rank well the seminal nodes and the level of agreement between the metric's biases and the biases implicitly present in the chosen set of seminal nodes. Note that until now, we discussed specifically the age bias because it is both manifestly present as well as easy to define and measure. Other potentially relevant biases---such as the field bias, for example---can be in principle studied and treated in a similar way as we do here for the age bias.

\section{Metric performance in ranking the seminal nodes whilst penalizing biased metrics (Task 2)}
\label{sec:task2}
Having demonstrated the caveats of evaluating the ranking performance of metrics using identification rate, we now proceed to Task 2 that additionally penalizes biased metrics. To this end, we employ the \emph{normalized identification rate} introduced in \citet{mariani2016identification} which imposes a penalty on metrics that are biased.

\subsection{Normalized identification rate}
\label{sec:NIR}
Normalized identification rate ($NIR$) introduced in \citet{mariani2016identification} considers the age distribution of the top-ranked nodes and applies a penalty factor to the identified seminal nodes that come from age groups that are over-represented among the top-ranked nodes. To compute $NIR$, we divide all $N$ network nodes by age into $G$ groups of equal size, and compute $N_z(g)$ which is the number of nodes from each group $g$ ($g=1,\dots,G$) that are in the top $z$ fraction of the ranking. An age-unbiased metric would result in $N_U:=zN/G$ top nodes, on average, in each age group. For any age group $g$ that is ``over-represented'' (that is, $N_z(g) > N_U$), the seminal nodes that are in the top $z$ fraction of the ranking do not contribute to the $NIR$ fully but only proportionally to $N_U / N_z(g)$. If, for example, a seminal node is from an age group that is \emph{twice} as frequent in the top of the ranking as it should be, this seminal node contributes only \emph{half} to the $NIR$. By contrast, seminal nodes from under-represented age groups ($N_z(g) < N_U$) contribute to the $NIR$ in the same way as they contribute to the $IR$. In other words, $NIR$ assumes a penalty for seminal nodes from over-represented age groups but no bonus for seminal nodes from under-represented age groups. To summarize, the factor $N_U/N_z(g)$ introduced by the normalized identification rate can be viewed as a penalty for the performance gained by age bias of a metric.

The choice of the number of age groups $G$ used in the computation of $NIR$ is a compromise between improving the temporal resolution (lowering the time duration of each group) by increasing $G$ and limiting the natural statistical variability of $N_z(g)$ by keeping $G$ low. We use $G=40$ adopted in previous literature \citep{mariani2016identification,vaccario2017quantifying}; other choices lead to qualitatively similar results. Note that due to the introduction of a penalizing factor, $NIR$ cannot be higher than $IR$ for a given ranking. The highest possible $NIR$ of one is achieved by a ranking that places all seminal nodes in the chosen top fraction of the ranking (we use top 1\% here, unless stated otherwise) \emph{and} the ranking is not biased by node age (or, at least, the age bins containing the seminal nodes are not over-represented in the top of the ranking).

The right panel of Figure~\ref{fig:hep_IR_NIR} shows the normalized identification rate as a function of the seminal node age for a small number of selected metrics. It shows that using $NIR$ indeed solves the problem encountered when metric performance is measured using the ordinary identification rate (left panel in Figure~\ref{fig:hep_IR_NIR}). In particular, AgeR becomes the worst method regardless of the seminal node age, as appropriate for a ranking method that actually ignores node impact in the network. LeaderRank, another metric that is strongly biased towards old nodes, is also strongly affected and its performance starts to decrease at the seminal node age 10 years instead of growing monotonously when identification rate is used. This illustrates that the use of the normalized identification rate weakens the mutually reinforcing link between age-biased metrics and age-biased sets of seminal nodes.

\begin{figure}
\centering
\includegraphics[scale=0.25]{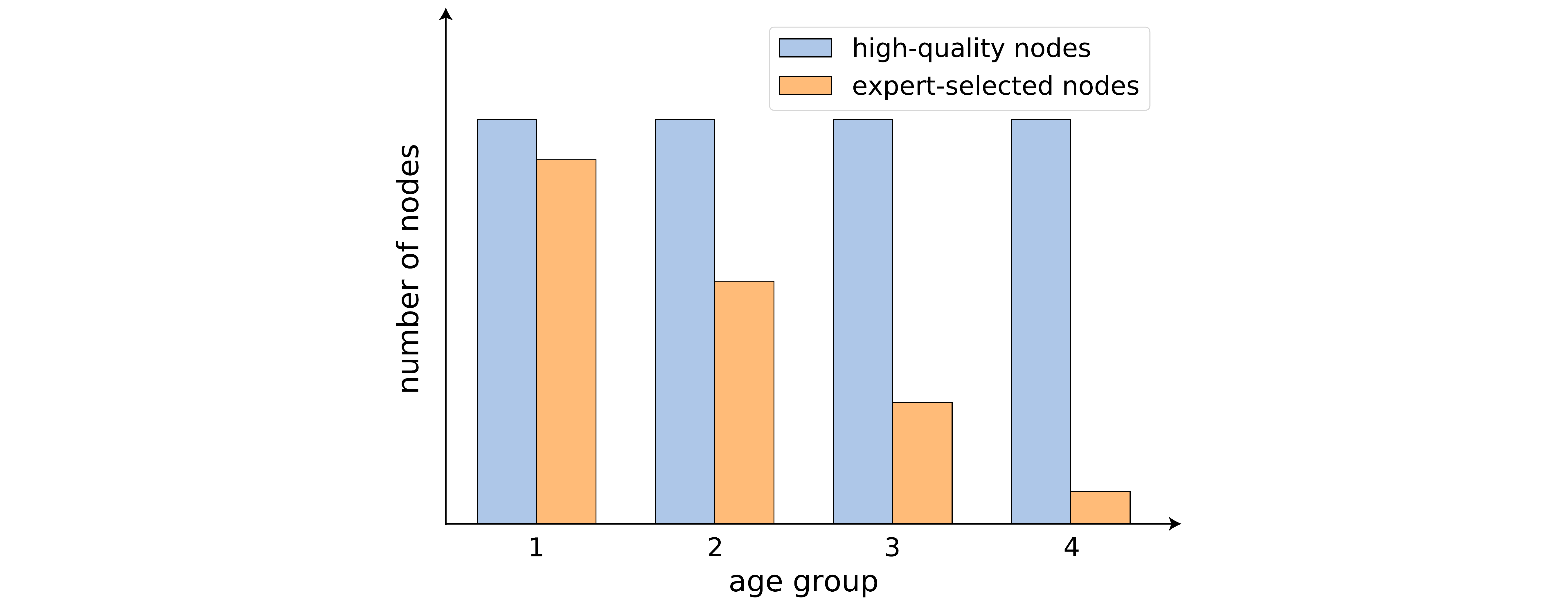}
\caption{An illustration for the alternative interpretation of Task 2: The age distribution of the seminal expert-selected nodes and the top $z$ fraction of nodes from each age group (we use here four age groups as an example).}
\label{fig:relationship_twonodes}
\end{figure}

There is also an alternative view that leads to Task 2 and the normalized identification rate as the appropriate evaluation methods. This view is based on realizing that the seminal nodes are necessarily a subset of high-quality nodes---the proverbial ``tip of an iceberg''. The short text introducing the Physical Review Letters milestones explicitly acknowledges that ``It is inevitable that some very important work will not be featured (in the milestones collection)''. We define the best $z$ fraction of nodes in each age group as the high-quality nodes---but the problem is that we do not know which are those ``best'' nodes. That is why we still need the seminal nodes but we do not view them as a definite and only target for the evaluated ranking metrics (which would correspond to Task 1). Instead, we recognize that the seminal nodes are a particular sample from all high-quality nodes in the dataset. This is illustrated by Figure~\ref{fig:relationship_twonodes} where the number of seminal nodes varies greatly among the age groups but the number of the top $z$ fraction of nodes is naturally constant. If a ranking metric over-represents a certain age group in its top $z$ fraction of the ranking by factor $X>1$, only the fraction $1/X$ of the top nodes from this age group are among the top $z$ fraction of nodes from this age group. That is why the factor $1/X$ needs to be applied to any identified seminal nodes from this age group---which is precisely what the normalized identification rate does. In summary, Task 2 and the normalized identification rate correspond also to the task of ranking highly the best nodes from each age group, from which the given seminal nodes are a potentially biased sample.

\begin{figure}
\centering
\includegraphics[width=\textwidth]{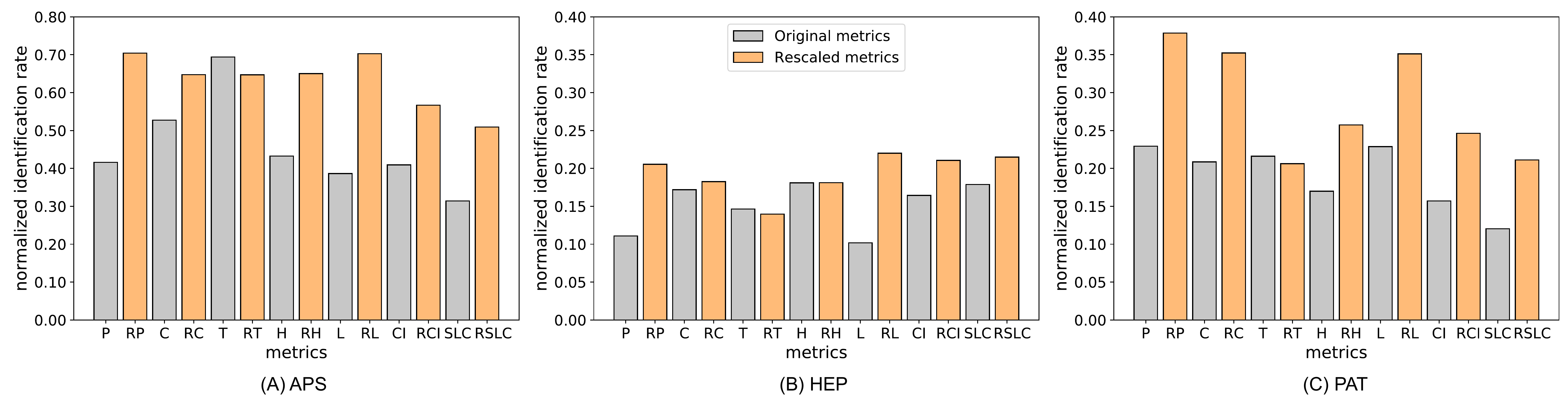}
\caption{Metrics' performance in identifying the seminal nodes as measured by $NIR$ evaluated on the complete datasets.}
\label{fig:performance_NIR}
\end{figure}

\subsection{Metric evaluation using normalized identification rate}
$NIR$ values on the complete datasets are shown in Figure~\ref{fig:performance_NIR}. The first thing to note is that the rescaled metrics generally perform better than their original counterparts here. The only exception is $T$ which itself contains a mechanism to prevent the metric from being overly biased towards old nodes, so an additional rescaling procedure is in some sense superfluous (even when \citet{mariani2016identification} and Figures \ref{SIfig:rankingbias_APS}, \ref{SIfig:rankingbias_HEP} and \ref{SIfig:rankingbias_PAT} show that CiteRank still displays strong age bias in some age groups). The second observation is that the $NIR$ values for the HEP dataset are much lower than the previously reported $IR$ values. This is a direct consequence of the penalization introduced by $NIR$ that heavily penalizes biased ranking metrics and, at the same time, unbiased ranking metrics being unable to well identify the biased seminal nodes. In line with the identification rate results in Task~1, YCCP performs similarly to RC: its $NIR$ values are 0.678 (APS), 0.175 (HEP), and 0.348 (PAT); the average score reported for the other metrics in Table~\ref{tab:summary} is 0.89.

\begin{table}
{\scriptsize\centering
\begin{tabular}{rrrrrrrrrrrrrrrrrr}
\toprule
   metric &   RP &   RL &   RC &   RH &  RCI & RSLC &    T &   RT &    C &    H &   CI &    P &    L &  SLC & HITS & RHITS\\
\midrule
avg score & 0.98 & 0.98 & 0.89 & 0.81 & 0.80 & 0.75 & 0.74 & 0.70 & 0.69 & 0.63 & 0.58 & 0.57 & 0.54 & 0.53 & 0.27 & 0.25\\
 rank APS &    1 &    2 &    5 &    4 &    7 &    9 &    3 &    6 &    8 &  10  &   12 &   11 &   13 &   14 &   15 &   16\\
 rank HEP &    4 &    1 &    5 &    6 &    3 &    2 &   11 &   12 &    9 &   7  &   10 &   15 &   16 &    8 &   14 &   13\\
 rank PAT &    1 &    3 &    2 &    4 &    5 &    9 &    8 &   11 &   10 &  12  &   13 &    6 &    7 &   14 &   15 &   16\\
\midrule
 avg score$_{z=0.005}$  & 0.96 & 0.96 & 0.78 & 0.73 & 0.72 & 0.68 & 0.65 & 0.63 & 0.60 & 0.65 & 0.53 & 0.55 & 0.49 & 0.46 & 0.20 & 0.21\\
 avg score$_{z=0.02}$   & 0.93 & 0.94 & 0.89 & 0.81 & 0.82 & 0.71 & 0.71 & 0.72 & 0.66 & 0.62 & 0.59 & 0.56 & 0.52 & 0.52 & 0.28 & 0.24\\
\bottomrule
\end{tabular}}
\caption{A summary of the metrics evaluation by normalized identication rate. The average score of metric $m$ is obtained by computing its score relative to the best-performing metric in each dataset, $NIR(m)/\max_n NIR(n)$, and averaging this score over the three analyzed datasets. A metric that would perform best in all datasets would therefore achieve the average score of one. The subsequent rows show the ranking of metrics by their $NIR$ for each dataset. The bottom part of the table shows the average score based on the $NIR$ values for two different top ranking fractions $z$, $0.5\%$ and $2\%$.}
\label{tab:summary}
\end{table}

The most important finding emerging from Figure~\ref{fig:performance_NIR} is that upon adopting the normalized identification rate for the evaluation, metrics that perform well in all three datasets emerge. This is clearly visible in Table~\ref{tab:summary} where the relative performance of the ranking metrics in all three datasets is summarized. We see that rescaled PageRank, RP, and rescaled LeaderRank, RL, perform best or nearly-best in all three datasets (recall that LeaderRank is a modification of PageRank obtained by changing the teleportation term). This result is robust with respect to changing the ranking fraction $z$ that we use to evaluate the normalized identification rate: even when the relative order of some metrics changes, RP and RL remain the two best metrics by some margin. It is interesting to note here that while the two top metrics are global in the sense of taking the whole network structure into account, they are followed directly by rescaled citation count, RC, which is a local metric that is based only on the immediate node neighborhood. Semi-local metrics, such as the semi-local centrality SLC and the collective influence CI, regardless if rescaled or not, combine the worst of both worlds: They are computationally more demanding than local metrics, and they rank nodes worse than local metrics.

Further insights can be gained by plotting $NIR$ as a function of the seminal node age in Figure~\ref{fig:relative_result_NIR} similarly as we did for the identification rate, $IR$, in Figure~\ref{fig:relative_result}. We see there, for example, that unlike in the APS data, rescaled PageRank does not outperform rescaled citation count in the HEP data in the first 18 years of seminal node age.

\section{Discussion}
\label{sec:discussion}
Previously introduced normalized identification rate ($NIR$) takes into account both the ranking positions of expert-selected nodes as well as the metric's bias that manifests itself in the ranking. We use $NIR$ to uncover the consistent performance of impact ranking metrics across different citation datasets. Our results indicate that ranking based on the network structure is more successful than simple degree-based metrics in singling out the significant nodes.

\subsection{Limitations and open directions}
There are various questions that remain open for further research. First, to extend our analysis to more than one kind of bias (such as the age and field bias common in scholarly citation data~\citep{vaccario2017quantifying}) and to generalize it to cases where the kind of bias is not explicitly known. Second, identification rate (referred to as recall in information filtering and statistical learning) is just one of various performance metrics (see \citep{dunaiski2018evaluate} for an overview of possibilities); we thus need to study how to account for bias in these other metrics. Third, besides metric evaluation on real data, evaluation on synthetic model data can be used to gain further theoretical insights. This approach has the advantage of having the possibility to arbitrarily turn on and off various model components and thus identify which of them are crucial for the observed metric performance. For example, which assumptions about papers written by multiple authors must be fulfilled in order for a specific researcher-impact metric reflect well the individual authors' contributions? Such models need to be informed by analyses of real empirical datasets and, conversely, the assumptions and effects identified as crucial in model data can be in turn validated in real datasets. Finally, there are other citation datasets that can be used for a similar analysis upon identifying corresponding sets of seminal nodes for them.

To draw a parallel, in a machine learning problem, if the training set has some intrinsic bias, the system learns this bias and in turn produces biased outcomes~\citep{raghavendra2018data,lloyd2018bias}. This is similar to our situation where a potential bias in the used set of seminal nodes, if left unchecked, can lead to wrong ranking metrics being believed to perform best, or even designing new ranking metrics that perform well only thanks to the bias. Biased outcomes of those metrics can in turn misguide our future evaluations and decisions. Further research of various aspects of bias in data mining and complex systems research, in particular how to avoid it, is therefore vital.

\subsection{Management implications}
Ranking and prioritizing is an essential task in many managerial applications. Our results show that to rank papers, age-rescaled PageRank is a well-performing metric that by construction produces rankings with little residual age bias. Evaluation of researchers and institutes typically uses metrics derived from citation counts (such as the $h$-index, for example). Our analysis, in particular the performance gap between age-rescaled PageRank and the tested unbiased versions of citation count, suggest that applying structural network-based metrics such as PageRank might be of advantage also when the objective is to rank the researchers or institutions.

\section{Conclusion}
Well-designed robust evaluation protocols are crucial for understanding which ranking metrics, which we have abundance of~\citep{waltman2016review,liao2017ranking}, perform well in which contexts. This study shows that the evaluation of a ranking metric based solely on the positions of expert-selected nodes in the resulting ranking is difficult to interpret because it confounds two aspects: the metric's ranking performance and the degree to which the biases of the expert-selected nodes overlap with the metric's biases. Normalized identification rate weakens the link between the ranking bias and the evaluation results, and yields results that are consistent across different datasets. In our case of ranking seminal nodes in citation networks, we find that age-rescaled PageRank and age-rescaled LeaderRank (note that LeaderRank is a close variant of PageRank) are the two best-performing metrics by a wide margin.

Our work deepens the understanding of impact metrics, especially in relation to the interplay between their biases and the biases of the considered test set. The comprehensive comparisons among various metrics are crucial to cope with the ever-growing number of new metrics and beneficial to understand the advantages and limitations of each of them. The proposed evaluation framework which penalizes biased metrics has general applicability beyond the ranking of articles in citation data; by highlighting the various roles of bias, it provides practical lessons for the ranking practice in other datasets with bias, such as technological networks, social networks, and other systems.

\section*{Acknowledgements}
This work is supported by the National Natural Science Foundation of China (Nos. 11622538, 61673150, 11850410444), the Science Strength Promotion Program of the UESTC, and the Zhejiang Provincial Natural Science Foundation of China (Grant no. LR16A050001). MSM acknowledges financial support from the University of Zurich through the URPP Social Networks, the Swiss National Science Foundation (Grant No. 200021-182659), the UESTC professor research start-up (Grant No. ZYGX2018KYQD21).

\section[*]{References}
\bibliographystyle{plainnat}
\bibliography{bibtex}

\appendix
\section{Evaluation of the age bias removal}
\setcounter{figure}{0}
\setcounter{table}{0}
\label{sec:bias}

\begin{figure}
\centering
\includegraphics[width=\textwidth]{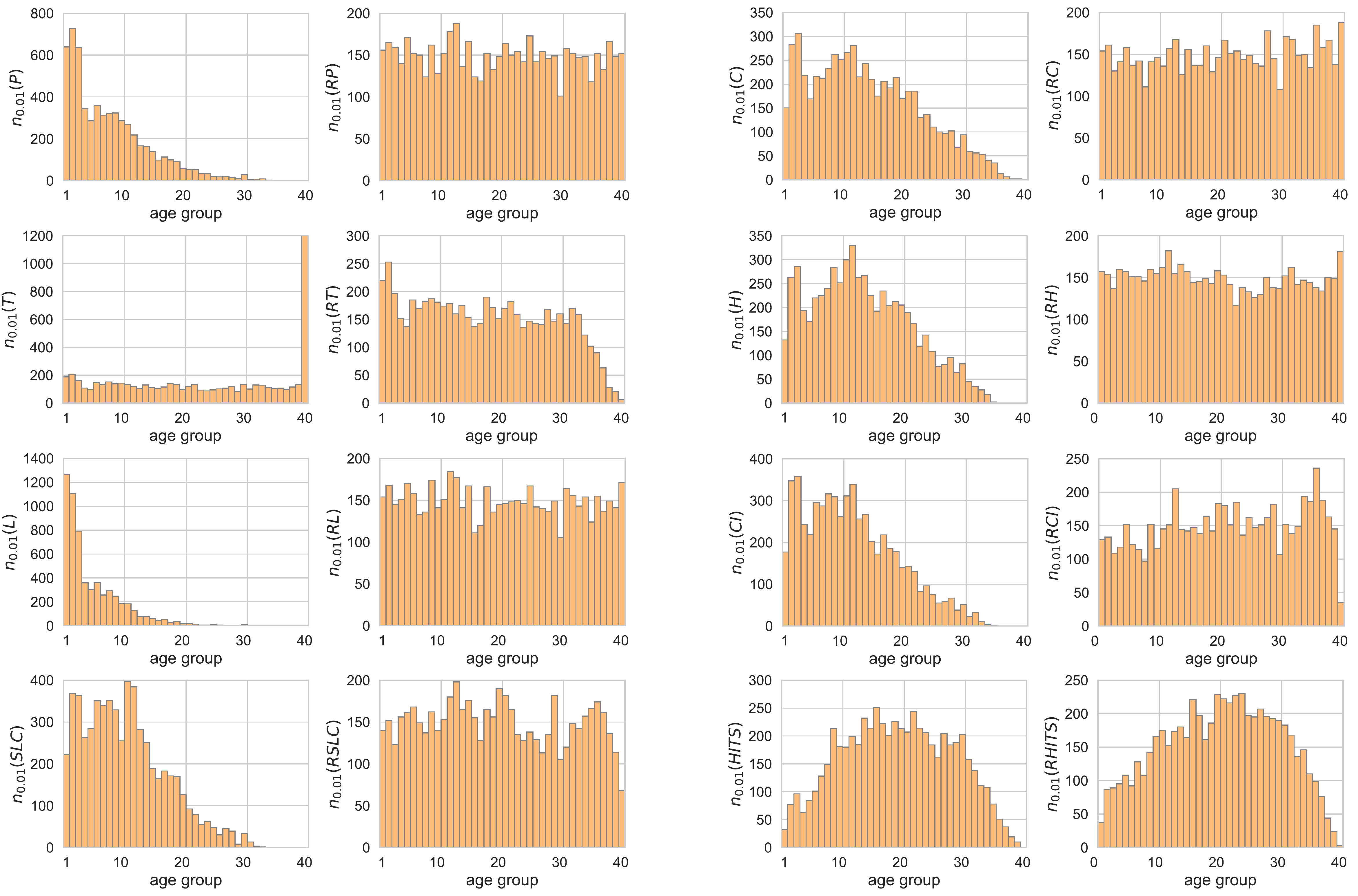}
\caption{Visualization of the age bias of original and rescaled metrics for the APS data.}
\label{SIfig:rankingbias_APS}
\end{figure}

\begin{figure}
\centering
\includegraphics[width=\textwidth]{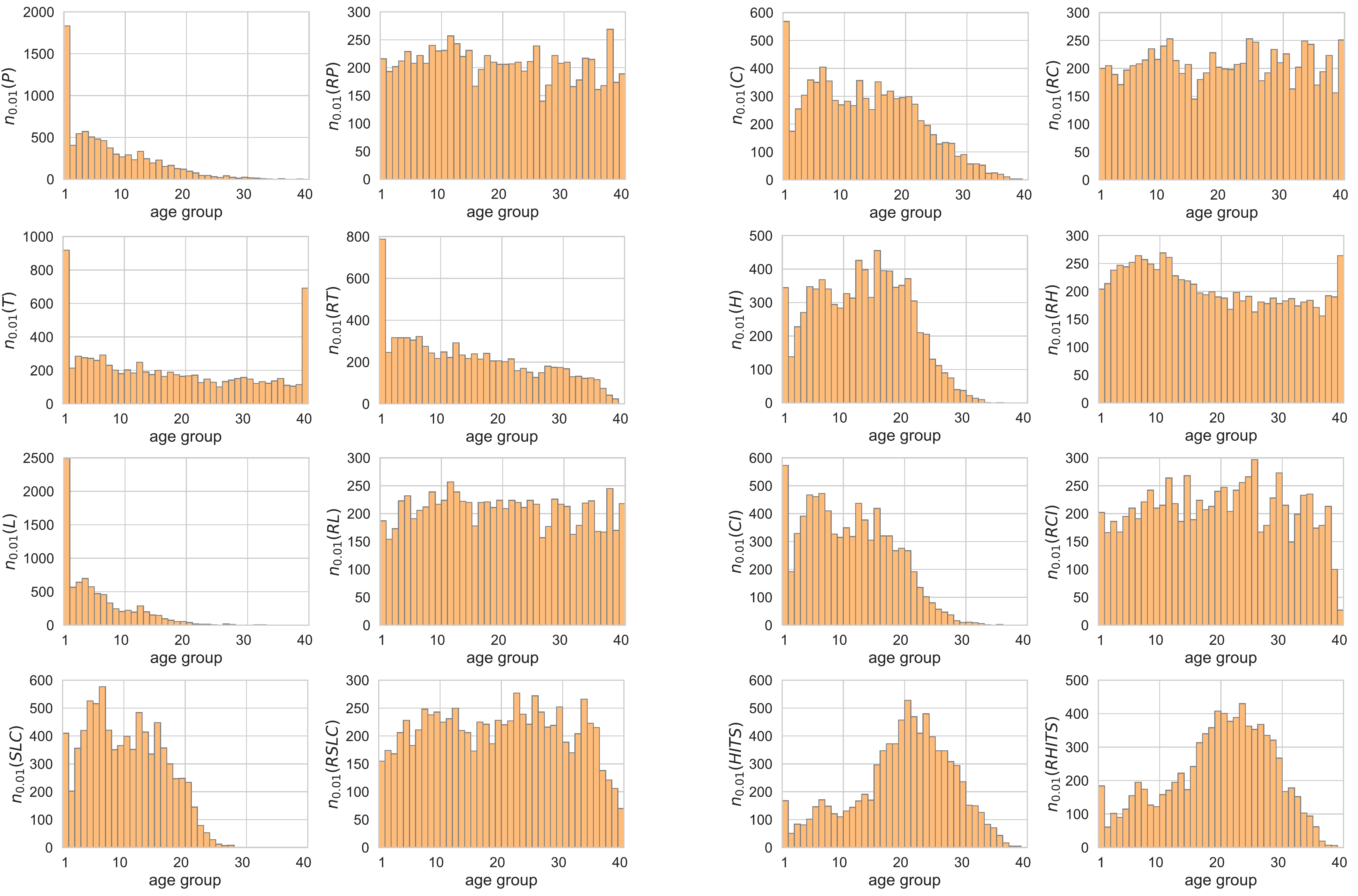}
\caption{Visualization of the age bias of original and rescaled metrics for the HEP data.}
\label{SIfig:rankingbias_HEP}
\end{figure}

\begin{figure}
\centering
\includegraphics[width=\textwidth]{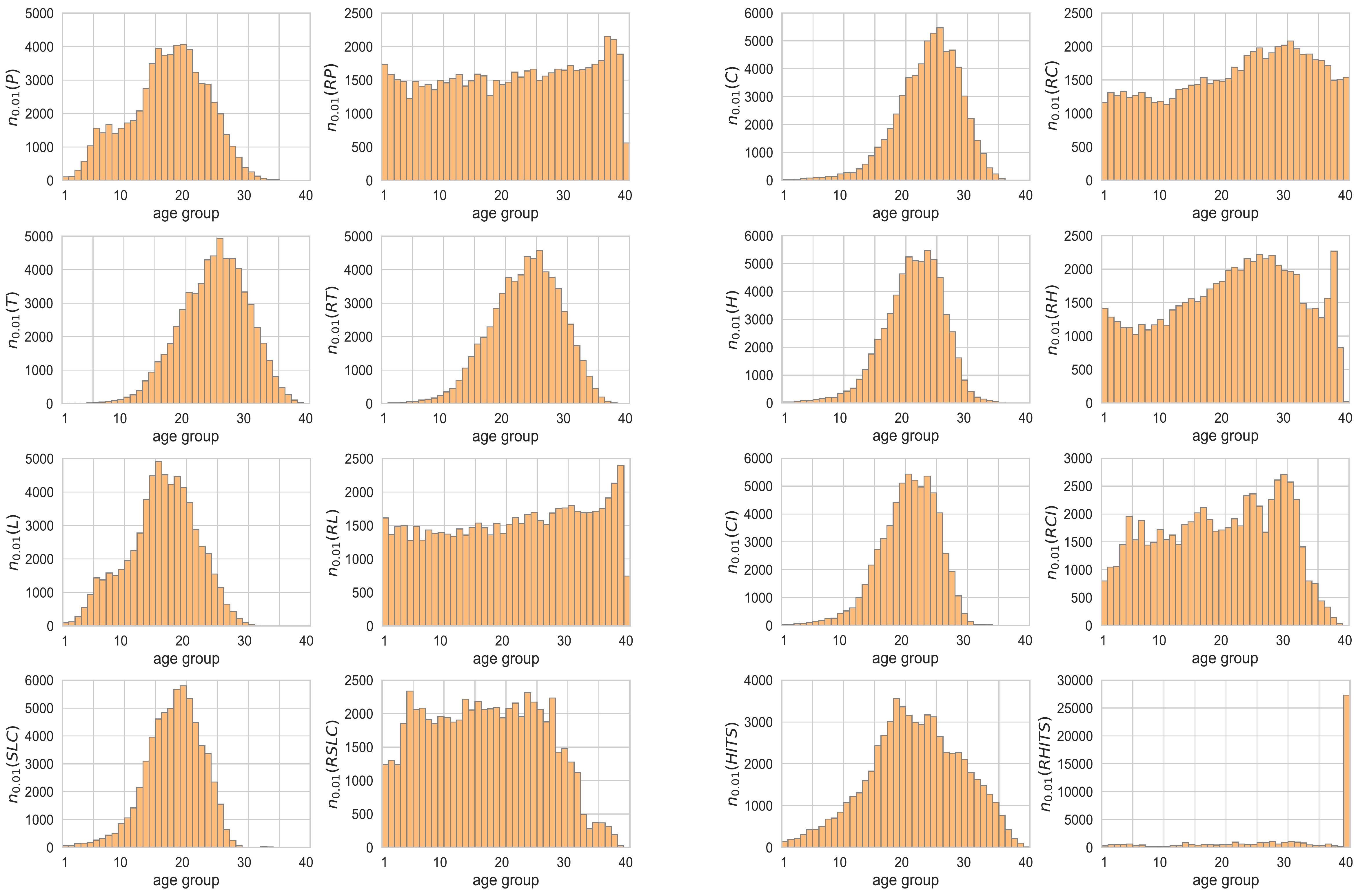}
\caption{Visualization of the age bias of original and rescaled metrics for the PAT data. The failure of weakening the age bias of the HITS authority score by rescaling is due to the authority scores ``concentrating'' on a small fraction of nodes with the remaining nodes having asymptotically zero score, which poses obvious problems to the rescaling procedure based on computing score mean and standard deviation in a finite moving time window.}
\label{SIfig:rankingbias_PAT}
\end{figure}

Figures~\ref{SIfig:rankingbias_APS}--\ref{SIfig:rankingbias_PAT} review the age bias of individual ranking metrics in the three analyzed datasets. Using the usual division of all nodes in 40 equally-sized groups by age, the figures show the number of nodes from each age group, $N_{1\%}(g)$ in the top 1\% of the ranking by each respective ranking metric. An age-unbiased metric would thus display a flat histogram where deviations from the perfectly uniform value $N_U=0.01N/40$ in each age bin would be only of statistical nature. As in \citep{mariani2016identification},
we measure the level of age bias in each histogram using the observed standard deviation
$$
\sigma = \sqrt{\frac1{40}\sum_{g=1}^{40} \big[N_{1\%}(g) - N_U\big]^2}
$$
with the average standard deviation $\sigma_0$ that results from distributing $0.01N$ nodes among the 40 age bins in a random (and therefore unbiased) way. When the bias strength is measured by $\sigma/\sigma_0$, the value of around one (or less) indicates that the observed level of bias can be explained by statistical fluctuations only. The higher the value, the stronger the bias. The values of $\sigma/\sigma_0$ corresponding to the histograms in Figures~\ref{SIfig:rankingbias_APS}--\ref{SIfig:rankingbias_PAT} are summarized in Table \ref{tab:bias_strength}.

\begin{table}
\centering
\begin{tabular}{rrrrrrr}
\toprule
& \multicolumn{2}{c}{APS} & \multicolumn{2}{c}{HEP} & \multicolumn{2}{c}{PAT}\\
         \cmidrule(r){2-3}         \cmidrule(lr){4-5}        \cmidrule(l){6-7}
metric & original & rescaled & original & rescaled & original & rescaled\\
\midrule
P    & 15.7 & 1.4 & 22.1 & 1.9 & 36.1 &  6.2\\
C    &  7.5 & 1.4 &  9.7 & 1.9 & 46.6 &  7.4\\
T    & 14.2 & 4.1 & 10.5 & 8.6 & 42.0 & 40.7\\
H    &  8.4 & 1.1 & 11.0 & 2.3 & 48.8 & 11.9\\
L    & 23.7 & 1.4 & 29.5 & 1.8 & 41.3 &  6.6\\
CI   &  9.9 & 2.8 & 12.5 & 3.3 & 49.6 & 17.9\\
SLC  & 11.5 & 2.1 & 13.8 & 3.1 & 50.6 & 18.6\\
HITS &  6.0 & 5.0 & 10.4 & 8.9 & 28.2 & 108.3\\
\bottomrule
\end{tabular}
\caption{Quantification of the age bias magnitude of respective ranking metrics in Figures \ref{SIfig:rankingbias_APS}, \ref{SIfig:rankingbias_HEP} and \ref{SIfig:rankingbias_PAT} using $\sigma/\sigma_0$ (the higher the value, the stronger the age bias; the value of one indicates zero bias).}
\label{tab:bias_strength}
\end{table}

\begin{figure}
\centering
\includegraphics[width=\textwidth]{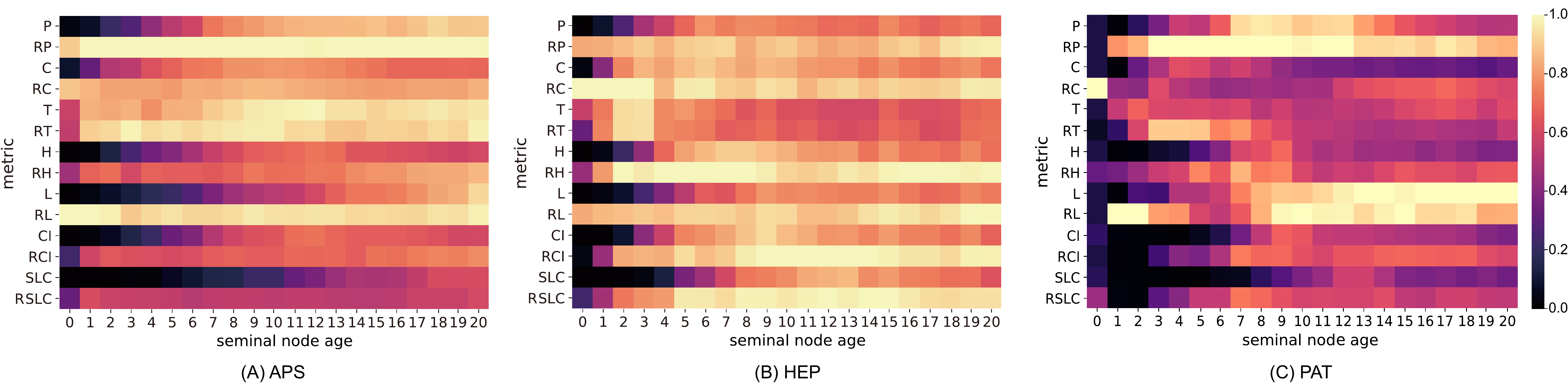}
\caption{The normalized identification rate of individual metrics as a function of the seminal node age (in years). To facilitate the comparison, the metrics' performance is normalized to the best metric in each age bin in the same way as in Figure~\ref{fig:relative_result}. A metric with zero $NIR$ thus receives zero score, while a metric that achieves the best $NIR$ for given seminal node age receives the score of one. }
\label{fig:relative_result_NIR}
\end{figure}

\end{document}